\documentclass[sigconf,authorversion,nonacm]{acmart}
\AtBeginDocument{%
  }

\setcopyright{cc}
\copyrightyear{2025}
\acmYear{2025}
\begin{document}

%%
%% The "title" command has an optional parameter,
%% allowing the author to define a "short title" to be used in page headers.
\title{AI Solutionism and Digital Self-Tracking with Wearables}

%%
%% The "author" command and its associated commands are used to define
%% the authors and their affiliations.
%% Of note is the shared affiliation of the first two authors, and the
%% "authornote" and "authornotemark" commands
%% used to denote shared contribution to the research.
\author{Hannah R. Nolasco}
\email{df104001@st.osakafu-u.ac.jp}
\orcid{0001-5801-8850}
\affiliation{%
  \institution{Osaka Prefecture University}
  \city{Sakai}
  \state{Osaka}
  \country{Japan}
}

\author{Andrew Vargo}
\affiliation{%
  \institution{Osaka Metropolitan University}
  \city{Sakai}
  \country{Japan}}
\email{awv@omu.ac.jp}

\author{Koichi Kise}
\affiliation{%
  \institution{Osaka Metropolitan University}
  \city{Sakai}
  \country{Japan}}
\email{kise@omu.ac.jp}

%%
%% By default, the full list of authors will be used in the page
%% headers. Often, this list is too long, and will overlap
%% other information printed in the page headers. This command allows
%% the author to define a more concise list
%% of authors' names for this purpose.
\renewcommand{\shortauthors}{Nolasco et al.}

%%
%% The abstract is a short summary of the work to be presented in the
%% article.
\begin{abstract}
Self-tracking technologies and wearables automate the process of data collection and insight generation with the support of artificial intelligence systems, with many emerging studies exploring ways to evolve these features further through large-language models (LLMs). This is done with the intent to reduce capture burden and the cognitive stress of health-based decision making, but studies neglect to consider how automation has stymied the agency and independent reflection of users of self-tracking interventions. In this position paper, we explore the consequences of automation in self-tracking by relating it to our experiences with investigating the Oura Ring, a sleep wearable, and navigate potential remedies.\footnote{This work was presented at the ACM CHI Conference on Human Factors in Computing Systems (CHI ’25) Workshop on Resisting AI Solutionism: Where Do We Go from Here? Yokohama, Japan.}
\end{abstract}

%%
%% The code below is generated by the tool at http://dl.acm.org/ccs.cfm.
%% Please copy and paste the code instead of the example below.
%%
\begin{CCSXML}
<ccs2012>
   <concept>
       <concept_id>10003120.10003121.10003126</concept_id>
       <concept_desc>Human-centered computing~HCI theory, concepts and models</concept_desc>
       <concept_significance>500</concept_significance>
       </concept>
 </ccs2012>
\end{CCSXML}

\ccsdesc[500]{Human-centered computing~HCI theory, concepts and models}

%%
%% Keywords. The author(s) should pick words that accurately describe
%% the work being presented. Separate the keywords with commas.
\keywords{Artificial Intelligence, AI Solutionism, Digital Health, Self-Tracking, Wearables, Sleep}

%\received{09 March 2025}
%\received[revised]{12 March 2009}
%\received[accepted]{5 June 2009}

%%
%% This command processes the author and affiliation and title
%% information and builds the first part of the formatted document.
\maketitle

\section{Introduction}

The Quantified Self (QS) movement encourages a systematic practice of self-tracking that is mediated by technology, meant to augment the act by allowing one to capture large amounts of personal data at a much faster rate and with greater accuracy~\cite{feng_how_2021}. This is supposed to empower one into optimizing their lifestyle for the purpose of improving their wellbeing or productivity~\cite{choe_understanding_2014, li_stage-based_2010, ruckenstein_visualized_2014} while foisting the burden of data collection from the user to the machine~\cite{choe_semi-automated_2017}. Self-tracking facilitated through wearable devices can collect physiological data passively, and through the accurate capture of physiological health markers, the device can aggregate insights often with the support of embedded artificial intelligence systems~\cite{wu_scientometric_2020}. This automation in the capture of personal data and in the dispensing of insights is perceived as a benefit to the user, given it reduces the cognitive load of having to determine the actions required to improve one’s health as captured by the technology~\cite{choe_semi-automated_2017}. Despite this, researchers remain divided on the impact of digital self-tracking and wearable devices—although there are studies which report an increase in decision making capabilities and self-confidence in users, others report experiences of anxiety and self-doubt, a loss of agency, and alienation from one’s perception of the self~\cite{wieczorek_ethics_2023}.

There is no evidence to support the notion that automation in self-tracking of health parameters encourages behavior change~\cite{patel_wearable_2015}. Yet most self-tracking tools and devices continue to be designed in such a way that curtails the involvement of the user in the process. With Large Language Models (LLMs) now coming into vogue, a growing body of literature is emerging which explores the implementation of LLMs into wearables to further automate the processing of device feedback~\cite{ahmed_leveraging_2025}, health monitoring and behavior modeling~\cite{ferrara_large_2024, kim_health-llm_2024}, and activity recognition~\cite{ferrara_large_2024, imran_llasa_2024}. In this position paper, we discuss the myriad hurdles to leveraging digital self-tracking interventions posed by automation and how it affected participants of our ongoing investigation of the Oura Ring, a sleep monitoring wearable. We contemplate upon the degrees of automation present in these technologies, which risk rendering users impotent to form independent decisions and embodied perceptions, and list ways the development of self-tracking interventions can be steered toward a more user-centered approach.

\section{Hurdles to Leveraging Automated Self-Tracking}

%This section underscores the prevailing issues faced by users of digital self-tracking interventions as a result of the tool or device’s effort to automate facets of the self-tracking process. We use these issues to frame the experiences of the participants in our ongoing investigation of the Oura Ring, a sleep monitoring wearable.

\subsection{Impairment to Information Literacy and Perception of Device Impact}

Commercially ready wearable sleep-trackers are equipped with sensors that are comparable to research grade equipment and are effective at tracking and assessing substantial amounts of personal data~\cite{de_zambotti_wearable_2019}. The Oura ring is one such device evidenced to be accurate in its sleep stage detection~\cite{altini_promise_2021} and its prediction of sleep quality~\cite{malakhatka_monitoring_2024}. Despite its sensor fidelity, we encountered an absence of improvement in sleep in users of the ring after one year of consistent device adherence~\cite{nolasco_perception_2023}. The users exhibited stagnant trend lines in their objective data, which contradicted their own reports that the device had a positive impact on both their sleep and habits.

Wearable devices utilize feedback systems which make use of descriptive statistics, and the highly technical and context-agnostic nature of the communicated information often leads to disengagement, which in turn prevents users from complying with their device's suggestions~\cite{patel_wearable_2015, wulfovich_i_2019}. We suspect that this incongruence between the perceptions that users had of the device's impact and their objective data is due to Oura's identical data delivery methods. Users of self-tracking tools frequently struggle with contextualizing and interpreting the visualized information presented by their device or application~\cite{grammel_how_2010, rooksby_personal_2014} , and it is possible that the participants of our study failed to appraise their data and thus were unable to inform their habits in response to it; they merely formed an impression of wellbeing improvement conceivably due to a desire for the device to be effective. It is also possible that they did become more mindful of their sleep hygiene, but their efforts were not detected by the device.

\subsection{Outsourcing Reflection Leading to Alienation from Data}

The effort necessitated by wellness-based decision making ostensibly affects a person's ability to intervene in the aspects of their health which can be improved or optimized, and self-tracking tools seek to reduce this cognitive load by supplying users with generated insights derived by the technology~\cite{li_understanding_2011, cho_art_2021, choe_understanding_2017}. But without a vehicle for self-reflection or involvement in the data interpretation, users of self-tracking technologies become alienated from their data and become increasingly disengaged until they eschew attentiveness to the device altogether~\cite{li_understanding_2011}.

In a qualitative interview with a cohort of long-term users of the Oura Ring, we urged participants to share their usage habits and impressions of the wearable, their relationship with sleep and the value they attribute to it, and the degree in which they have absorbed the knowledge that the device should have conferred to them after several months of usage. Many of their sentiments expressed a dissatisfaction with the information gained from the ring, as they felt it did not tell them anything they did not already know, or it contradicted their personal evaluations of their own condition. This aligns with the experiences of other users of digital self-tracking, who revealed that the data automatically captured by the device does not abide by what interests them~\cite{lazar_why_2015} and that their objective data did not align with their own subjective observations~\cite{attig_abandonment_2020, rooksby_personal_2014, nolasco_augmenting_2024}.

Respondents of the interview also demonstrated a semblance of learned helplessness, in which they expressed beliefs of individual inadequacy to transform their lifestyle as a result of external factors outside of their control, such as school, work, and their social life. This indicates a potential need for users of the ring to develop more self-efficacy, or the confidence that they can successfully achieve a task or adopt new behavior. Self-efficacy is known to spur more engagement and adherence to device feedback~\cite{lin_continued_2019}, yet we suspect that this is impeded by the reliance that wearables can engender in its own users, which compels them to be dependent on its instruction~\cite{wieczorek_ethics_2023}.

\section{Discussion and Conclusion}

Digital self-tracking interventions through manual methods, in which the user participates in or is the sole driver of the tool’s data collection by means of their input, can positively impact users through behavioral change techniques (BCTs) which direct the user towards adopting better health-related behavior~\cite{oyebode_sleepfit_2021, choe_sleeptight_2015}. Wearable devices may also have BCTs embedded into the infrastructure of their interfaces, yet they seem unable to achieve the same impact~\cite{nolasco_perception_2023}, owing to the many obstacles that arise from automating the process of collecting and interpreting personal data. Automated data collection leads to limiting factors that makes the process inflexible to any forms of personalization, which only creates discrepancies between the data that can be captured and what the user is concerned about tracking~\cite{harrison_activity_2015, yang_when_2015}. It also confines the process to easily detectable phenomena (e.g., heart rate)~\cite{choe_semi-automated_2017}, without considering implicit factors like pain, social interactions, and emotions.

We have found that long-term users of the Oura Ring are unable to fully leverage the benefits of the device, which we believe is partly caused by the issues previously mentioned. The limitations in the types of data collected likely curbed their own interests and consequently stifled their engagement, which is known to compromise one's understanding of their wearable data~\cite{li_understanding_2011}. This may have led to a misplaced belief in the device’s impact despite a lack of positive change in their captured sleep quality~\cite{nolasco_perception_2023}, either due to inattention or misapprehension. Users of the ring also expressed feelings of disconnect from their data, a lack of information gain, and a sense of learned helplessness, which is potentially due to the generated insights provided by the device which prevents them from arriving at useful, actionable insight by themselves~\cite{choe_understanding_2017}. It comes as no surprise that automated self-trackers demonstrate the shallowest reflection compared to manual and mixed methods self-trackers~\cite{krukkert_manual_2023}.

\subsection{Future Work}

Several studies are scouring to prove that AI can extract exceedingly more accurate and granular information on the user's behalf while so few consider how users even absorb said information~\cite{baron_feeling_2018}. The obsession with automation assumes that the machine is superior in deciding what actions and habits will benefit us, be it due to what we imagine are unbiased calculations occurring unnoticed behind the backcloth or the conception that the machine can do perfectly what we can only do flawed. But this only makes us bystanders and altogether strangers to our own lived experiences. While manual self-tracking only offloads some elements of independent thought from the user, automated self-tracking offloads everything: from the data collection to the generation of insights. This not only stirs up preoccupations and anxieties over device appeasement~\cite{gutierrez_other_2016}, but also estranges the user from their data~\cite{gutierrez_other_2016,li_understanding_2011}, preventing them from recognizing what features of their daily life require their remediation. We must consider what value judgements we are making by automating the process of evaluating the self and giving machines the authority to define what a healthy person is, which commonly complies with the image of an "entrepreneurial citizen"~\cite{feng_how_2021} and not an inherently flawed body always in flux~\cite{sanders_self-tracking_2017}. Many of the health challenges faced by people are also the product of failures in state systems, which can only be mitigated through better policymaking and not by leaving it to private companies and their technologies to solve~\cite{gilmore_predicting_2021}.

Future directions in digital self-tracking should endeavor to separate the notion of ‘fast’ from ‘efficient’ and focus less on immediate and instant feedback, insight, and information. Instead, efforts should be made to pursue slower technology which merely encourages self-reflection without the pressure of appeasing a device or adhering to societal norms~\cite{hallnas_slow_2001}. People yearn for personalized, self-managed, non-automated tracking where the user has the power to define what matters~\cite{wannamaker_io_2021, patel_wearable_2015, wulfovich_i_2019}. But if automation is to be used, achieving a balance between manual and automated self-tracking is what best leverages the strengths of both methods~\cite{choe_semi-automated_2017, krukkert_manual_2023}.

\begin{acks}
This work was supported in part by grants from JST ASPIRE (JPMJAP2403), JSPS Grant-in-Aid for Scientific Research (23KK0188) and the Grand challenge of the Initiative for Life Design Innovation (iLDi).
\end{acks}

%%
%% The acknowledgments section is defined using the "acks" environment
%% (and NOT an unnumbered section). This ensures the proper
%% identification of the section in the article metadata, and the
%% consistent spelling of the heading.
%\begin{acks}
%To Robert, for the bagels and explaining CMYK and color spaces.
%\end{acks}

%%
%% The next two lines define the bibliography style to be used, and
%% the bibliography file.
\bibliographystyle{ACM-Reference-Format}
\bibliography{2025chi-w20}

%%
%% If your work has an appendix, this is the place to put it.
\appendix

\end{document}